\documentclass[twocolumn,aps,pra,superscriptaddress]{revtex4}

\usepackage{graphicx}
\usepackage{amsmath}
\usepackage{amssymb}
\usepackage[utf8]{inputenc}
\usepackage{dcolumn}
\usepackage{bm}
\usepackage{psfrag}
\usepackage{yfonts}
\usepackage{version}
\usepackage{natbib}
\usepackage{varioref}
\usepackage{units}
\usepackage{esint}
\usepackage{epsfig}
\usepackage[usenames,dvipsnames]{color}

\usepackage{dsfont}

\renewcommand{\exp}[1]{e^{#1}}

\newcommand{\bra}[1]{\langle\,{#1}\, |}
\newcommand{\ket}[1]{|\,{#1}\,\rangle}

\newcommand{\opP}[1]{\ket{\pi_{#1}}\bra{\pi_{#1}}}

\newcommand{\cm}{\ {\rm cm}^{-1}}

\usepackage{ulem}  
\newlength{\mylenunit}
\begin{document}

\title{
An efficient method to calculate excitation energy transfer in light harvesting systems. 
Application to the FMO complex.}

\author{Gerhard Ritschel}
\affiliation{Max-Planck-Institut f\"ur Physik komplexer Systeme, N\"othnitzer Str.~38, D-01187 Dresden, Germany}

\author{Jan Roden}
\affiliation{Max-Planck-Institut f\"ur Physik komplexer Systeme, N\"othnitzer Str.~38, D-01187 Dresden, Germany}

\author{Walter T.\ Strunz}
\affiliation{Institut f\"{u}r Theoretische Physik,
Technische Universit\"at Dresden, D-01062 Dresden, Germany}

\author{Alexander Eisfeld}
\email{eisfeld@mpipks-dresden.mpg.de}
\affiliation{Max-Planck-Institut f\"ur Physik komplexer Systeme, N\"othnitzer Str.~38, D-01187 Dresden, Germany}
\affiliation{Department of Chemistry and Chemical Biology
Harvard University
12 Oxford Street,
Cambridge, MA 02138}

\date{\today}

\begin{abstract}
A master equation, derived from the non-Markovian quantum state diffusion (NMQSD), is used to calculate excitation energy transfer in the photosynthetic Fenna-Matthews-Olson (FMO) pigment-protein complex at various temperatures.
This approach allows us to treat spectral densities that contain explicitly the coupling to internal vibrational modes of the chromophores.
Moreover, the method is very efficient, with the result that the transfer dynamics can be calculated within about one minute on a standard PC, making systematic investigations w.r.t.\ parameter variations tractable.
After demonstrating that our approach is able to reproduce the results of the numerically exact hierarchical equations of motion (HEOM) approach, we show how the inclusion of vibrational modes influences the transfer.

\end{abstract}


\maketitle

\section{Introduction}

Since the pioneering work of Franck and Teller~\cite{FrTe38_861_}, exciton theory is widely used to describe the optical and energy transfer properties of light harvesting systems~\cite{AmVaGr00__}. 
In recent years particular focus has been drawn to the so-called Fenna-Matthews-Olson (FMO) complex (see e.g.\ Refs.~\cite{MiBrGr10_257_,ChFl09_241_,ChVaBr05_10542_,ReMa98_4381_,BrMa04_10529_,RaeFr07_251_,BrStVa05_625_,EnCaRe07_782_,ReMoKa09_033003_,PaAbMu10_108_,MueMaAd07_16862_,ReMoAs09_9942_,MoReLl08_174106_,AdRe06_2778_,OlStSc11_758_,KrKrRo0__,WuLiSh10_105012_,SaChWh11_11906_,ARXIV_Sangwoo,BrEi11_051911_}), that promotes energy transfer from the main light harvesting complex  towards the reaction center in green sulfur bacteria~\cite{AmVaGr00__}.

\begin{figure}
\includegraphics[width=0.5\mylenunit]{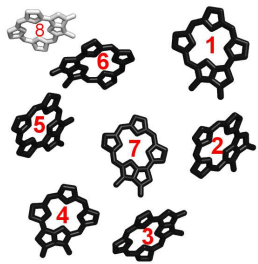}
\caption{\label{fig:FMO}
The numbering of the BChls within the FMO monomer. In the present work BChl 8
is not taken into account.
The figure is created using PyMOL~\cite{PyMOL}.
}
\end{figure}

The FMO complex consists of three identical subunits, called ``monomers'', each containing eight \cite{BeFrNe04_274_,TrWeGa09_79_} bacteriochlorophyll (BChl) molecules (see Fig.~\ref{fig:FMO}).
It is assumed that BChls~1, 6, and~8 are located at the baseplate and receive
electronic excitation captured by the chlorosomes \cite{AmMuMa11_93_}.   
BChls 3 and 4 are located in the vicinity of the reaction center.
To understand the excitation transfer mechanisms, many experimental and theoretical investigations, that took advantage of the availability of high resolution crystal structure, have been performed.
Supported by theoretical modeling a good overall understanding of the local energies of the chlorophyll molecules and their dipole-dipole couplings has been obtained, although there is still a large variance between  values found in the literature \cite{Unp_Olbrich}.

The present interest in the FMO complex has been greatly stimulated by the evidence for wavelike energy transfer through quantum coherence in photosynthetic systems \cite{EnCaRe07_782_}.
Since the electronic coherences are strongly influenced by the coupling of electronic excitation to vibrational modes of the BChl molecules and by the coupling to the protein environment, a detailed modeling of this interaction is important.
Information on this coupling has been obtained e.g.\ from  fluorescence line narrowing \cite{RaeFr07_251_,WePuPr00_5825_} or from theoretical simulations \cite{AdRe06_2778_,ARXIV_Sangwoo,Unp_Olbrich},
which indicate that the spectral density, describing the coupling of the electronic excitation to vibrational modes, is quite structured.
Such a structured spectral density will lead to non-Markovian effects.
The influence of these non-Markovian effects have been investigated by various theoretical methods \cite{ReChAs09_184102_,IsFl09_234111_,IsFl09_17255_,PrChHu10_050404_,TaMi10_891_,HuCo10_184108_,ARXIV_Nalbach}. 
In general these methods are difficult to apply to large systems and/or arbitrary spectral densities.

In the present paper we present an efficient approach, which allows us to treat structured spectral densities with minimal requirements on the computer capabilities. 
All the calculations presented below can be performed on a standard PC within a few minutes.
The method is based on the non-Markovian quantum state diffusion (NMQSD) approach \cite{DiSt97_569_,DiGiSt98_1699_,YuDiGi99_91_,StYu04_052115_,Yu04_062107_,VeAlGa05_124106_}  and allows us to treat the whole range from coherent wavelike motion to incoherent hopping.
 We have applied this method, within the zeroth order functional expansion (ZOFE) approximation, to describe the optical and transfer properties of molecular aggregates~\cite{RoEiWo09_058301_,RoStEi10_5060_,RoStEi11_034902_}. 
Within the ZOFE approximation it is possible to derive a non-Markovian master equation \cite{StYu04_052115_}, which we use in this paper (in the following we will call it ZOFE master equation).
This approach enables us to calculate the excitation dynamics of fairly large systems, e.g.\ aggregates consisting of more than $\sim\!30$ chromophores, taking structured spectral densities for the exciton-phonon coupling into account.

Although our method contains an approximation, from a previous study on absorption spectra of smaller complexes we expect that for parameters relevant for the FMO complex the approximation works  well \cite{RoStEi11_034902_}.
This will be confirmed in the following, where we will compare excitation energy transfer, calculated using the ZOFE master equation, with results of the so-called hierarchical equations of motion (HEOM) approach. 
 The HEOM approach has become popular \cite{XuCuLi05_041103_,IsFl09_234111_,StSc09_225101_,ChZhSh09_094502_,ZhKaRe11_1531_,KrKrRo0__},   since it allows in principle converged results to be obtained.
However, the numerical effort grows rapidly with increasing system size and for spectral densities that deviate from the Drude-Lorentz (D-L) form.
To overcome these difficulties extensive parallelization (e.g.\ by using graphics processing units \cite{KrKrRo0__}) is necessary.

Since in our method we are not restricted to a simple form of the  spectral density,  we can investigate how energy transfer depends on its details. 
In particular we show that the transfer
dynamics is {\it not} sensitive to coupling to modes with energies above the largest energy difference between the purely electronic eigenstates of the FMO complex (which is roughly $500 \cm$ ).
Furthermore, we demonstrate that the transfer dynamics depends on the details of the local structure of the spectral density in the region, where the FMO complex has electronic transition energies.

The paper is organized as follows:
In Section \ref{sec:Model} we introduce the model used to describe the FMO complex and we present the ZOFE master equation that we use for the numerical calculations.
In Section \ref{sec:num_cal} we first demonstrate that with the ZOFE master equation we are able to reproduce calculations obtained with the HEOM method. 
Then we study the influence of different parts of the spectral density.
Finally, in Section \ref{sec:summary} we conclude with a summary and an outlook.

\section{Model}
\label{sec:Model}
In the following we will briefly introduce the model Hamiltonian that we use to
describe the FMO complex. Further details can be found e.g.\ in Ref.~\cite{MaKue00__,RoStEi11_034902_}.

In the equations below, we set $\hbar=1$.
For each chromophore we take two electronic states into account. 
Since we are interested in excitation transfer of a single electronic
excitation, we restrict ourselves to the one-exciton subspace which is spanned
by the states
\begin{equation}
\ket{\pi_n}=\ket{g}\cdots \ket{e} \cdots \ket{g},
\end{equation}
where chromophore $n$ is excited electronically and the other BChls are in their electronic ground state. 
In this basis the electronic ``system'' part of the FMO complex is 
\begin{equation}
\label{H_sys}
  H_{\rm sys}=\sum_{n=1}^N\varepsilon_n\ket{\pi_n}\bra{\pi_n}+\sum_{n,m=1}^N V_{nm}\ket{\pi_n}\bra{\pi_m},
\end{equation}
where $\varepsilon_n$ are the electronic energies of the chromophores and 
$V_{nm}$ are the couplings between them.
 The ``environment'' of vibrational modes is taken in the form
\begin{equation}
\label{H_env}
  H_{\rm env}=\sum_{n=1}^N\sum_{\lambda}\omega_{n\lambda}a^{\dagger}_{n\lambda}a_{n\lambda}
\end{equation}
 and the coupling of electronic excitation to these vibrations is expressed through
\begin{equation}
\label{HInt}
  H_{\rm int}=-\sum_{n=1}^N\ket{\pi_n}\bra{\pi_n}\sum_{\lambda}\kappa_{n\lambda}(a^{\dagger}_{n\lambda}+a_{n\lambda}).
\end{equation}
Here the  $\kappa_{n\lambda}$, which describe the coupling of the electronic excitation on BChl $n$ to the local vibrational mode $\lambda$ at this chromophore,  are related to the spectral density $J(\omega)$ by \cite{MaKue00__}
\begin{equation}
\label{spec_dens}
J_n(\omega)=\sum_{\lambda}|\kappa_{n\lambda}|^2\ \delta(\omega-\omega_{n\lambda})
\end{equation}
which will be taken as a continuous function in the following.
For a thermal initial  state $\rho_{\rm th}$ of the environment the environment correlation function $\alpha_n(t-t')$ is related to the spectral density by the standard expression
\begin{eqnarray}
\label{bathCorrFunc}
\alpha_n(\tau)  =\int_0^{\infty}\! \!\!\! {\rm d}\omega \,J_n(\omega)\left[\coth\left(\frac{\omega}{2kT}\right)\cos(\omega\tau)-i\sin(\omega\tau)\right]
\end{eqnarray}
Note that we assume that the local environments of different chromophores are uncorrelated, which is in accordance with recent calculations \cite{OlStSc11_758_}. 
Finally the total Hamiltonian in the one-exciton space is given by
\begin{equation}
\label{HeAsSysPlusIntPlusEnv}
  H^e=H_{\rm sys}+H_{\rm int}+H_{\rm env}.
\end{equation}

In the following we consider  electronic excitation transfer.
Thus we will focus on the reduced density operator $\rho(t)$ in the electronic subspace, which is the trace of the total density operator $\rho_{\rm tot}(t)$ of system and environment over the vibrations
\begin{equation}
\rho(t)={\rm Tr}_{\rm env}\,\rho_{\rm tot}(t).
\end{equation}
Its diagonal elements in the $\ket{\pi_n}$ basis describe the time-dependent populations of the chromophores.

\subsection{Numerical method}

We calculate the reduced density operator by applying a recently developed method, which is based on the non-Markovian quantum state diffusion approach~\cite{DiGiSt98_1699_,RoEiWo09_058301_,RoStEi11_034902_,Yu04_062107_}.
Within the ZOFE approximation (described in detail in Ref.~\cite{YuDiGi99_91_,RoStEi11_034902_}) one can derive a convolutionless master equation for the reduced density operator \cite{StYu04_052115_}
\begin{align}
\label{zofe_mast_eq}
\partial_t\rho(t)=&- i[H_{\rm sys},\rho(t)]\nonumber\\
&-\sum_n\big[\opP{n},\rho(t)\,\bar{O}_0^{(n)\,\dagger}(t)\big]\\
&-\sum_n\big[\bar{O}_0^{(n)}(t)\,\rho(t), \opP{n}\big].\nonumber
\end{align} 
Here the operator
\begin{equation}
\label{DefOQuer}
  \bar{O}_0^{(n)}(t)=\int_0^t ds\ \alpha_n(t-s)O^{(n)}_0(t,s)
\end{equation} 
 is determined by the auxiliary equation   
\begin{align}
\label{eq:O_0}
  \partial_tO_0^{(n)}(t,s)=&-i\Big[H_{\rm sys},O^{(n)}_0(t,s)\Big]\\
&+\sum_m\Big[\opP{m}\bar{O}^{(m)}_0(t),O^{(n)}_0(t,s)\Big]\nonumber
\end{align}
with the initial condition $O^{(n)}_0(s,s)=-\opP{n}$ (we use the same notation as in Ref.~\cite{RoStEi11_034902_}  where the subscript 0 indicates that the operators are obtained from the ZOFE approximation).
For the numerical implementation, we approximate the environment correlation function as a sum of exponentials, i.e.\ $\alpha_n(\tau)=\sum_j^M p_{nj}\,\exp{ i \Omega_{nj}\tau}$ with in general complex prefactors $p_{nj}$ and complex frequencies $\Omega_{nj}$. 
Such a form allows us to derive a closed set of coupled  differential equations 
\begin{equation}
\begin{split}
 \partial_t\bar{O}_0^{(nj)}(t)=&-i\left[H_{\rm sys},\bar{O}_0^{(nj)}(t)\right]\\ 
&\ -p_{nj}\opP{n}+i\Omega_{nj}\bar{O}_0^{(nj)}(t)\\
&+\sum_m\left[\opP{m}\bar{O}^{(m)}_0(t),\bar{O}_0^{(nj)}(t)\right]
\end{split}
\end{equation}
for the  operators $\bar{O}^{(n)}_0(t)=\sum_j\bar{O}^{(nj)}_0(t)$ with initial condition $\bar{O}_0^{(nj)}(t=0)=0$.
In the numerical implementation we expand the equations above in the $\ket{\pi_n}$ basis.
To determine the $\Omega_{nj}$ we proceed similarly as in Ref.~\cite{MeTa99_3365_,KlScSc05_461_}.
However, in the present work we do not use the Matsubara decomposition for the $\coth$ appearing in the correlation function $\alpha_n(\tau)$ (see Eq.~(\ref{bathCorrFunc})) but use the partial fraction decomposition proposed by Croy and Saalmann \cite{CrSa09_073102_}. 
 The latter has superior convergence properties and uses poles in the complex plane (not only on the imaginary axis, as is the case for the Matsubara decomposition).
Since in the present approach one can use hundreds of exponentials, in principle it is no problem  to approximate the environment correlation function $\alpha(\tau)$ for  arbitrary spectral densities $J(\omega)$ and temperatures with high accuracy.

\section{Excitation energy transfer}
\label{sec:num_cal}

\subsection{Comparison with HEOM calculations}

First we will compare the energy transfer calculated  with the ZOFE master equation  approach with the HEOM results of Ref.~\cite{IsFl09_17255_}. 
For the comparison we restrict ourselves to the same model used in Ref.~\cite{IsFl09_17255_} where only a single monomer unit of the FMO trimer was considered and the recently discovered eighth BChl was not taken into account (see Hamiltonian in Table~\ref{tab:ham_mon}).
\begin{table}
\begin{tabular}{ c c c c c c c c }
\hline\hline
 & 1 & 2 & 3 & 4 & 5 & 6 & 7 \\ 
\hline\hline
1 & \textbf{410} & -87.7& 5.5& -5.9& 6.7& -13.7& -9.9\\
2 & & \textbf{530} & 30.8& 8.2& 0.7& 11.8&   4.3\\
3 & & & \textbf{210} & -53.5& -2.2& -9.6& 6.0\\
4 & & & & \textbf{320} & -70.7& -17.0& -63.3\\
5 & & & & & \textbf{480} & 81.1& -1.3\\
6 & & & & & & \textbf{630} & 39.7\\
7 & & & & & & & \textbf{440} \\
\hline\hline
\end{tabular}
\caption{Matrix elements of the Hamiltonian $H_{\rm sys}$ of the FMO monomer in $\cm$.
The upper triangle (couplings) is the same as the lower triangle (not shown here).
From the energies on the diagonal an offset of $12000\cm$ is subtracted (for the calculation of the transfer only the energy differences are relevant).
The Hamiltonian is that used in Ref.~\cite{IsFl09_17255_} with the energies and couplings from Ref.~\cite{AdRe06_2778_}.
}
\label{tab:ham_mon}
\end{table}

\begin{figure}
\includegraphics[width=0.85\mylenunit]{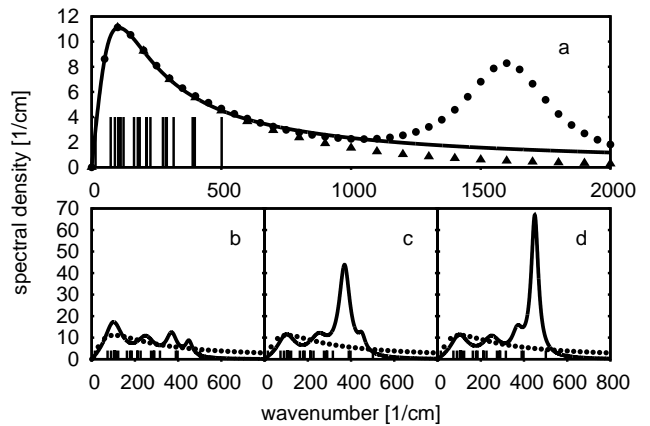}
\caption{\label{fig:spec_dens}
The different spectral densities considered.
(a) Solid line: D-L spectral density used in Ref.~\cite{IsFl09_17255_}. 
Triangles: as solid line, but with faster drop-off above $500\cm$.
Dots: same as solid line, but with extra peak at $1600\cm$.
(b)--(d)  Dots: as solid line in (a).
Solid line: fictitious structured spectral densities, where one peak at a time is enhanced, while the effective reorganization energy $\lambda_{\rm eff}$ is kept constant and equal to that of the dotted curve. 
The vertical lines indicate the values of all transition energies between the electronic eigenenergies of the FMO monomer unit.
} 
\end{figure}

For the comparison we have taken the transition energies and couplings from \cite{AdRe06_2778_}.
In Ref.~\cite{IsFl09_17255_} a Drude-Lorentz (D-L) spectral density was used to describe the coupling to the environment.
For the decomposition of the environment correlation function as a sum of exponentials, described above, we  represent the spectral density by a sum of anti-symmetrized Lorentzians.
With 10  anti-symmetrized Lorentzians we could fit the D-L spectral density of Ref.~\cite{IsFl09_17255_} perfectly in the relevant energy range. 
The resulting spectral density is shown as the solid curve in Fig.~\ref{fig:spec_dens}(a). 

\begin{figure} 
\includegraphics[width=0.85\mylenunit]{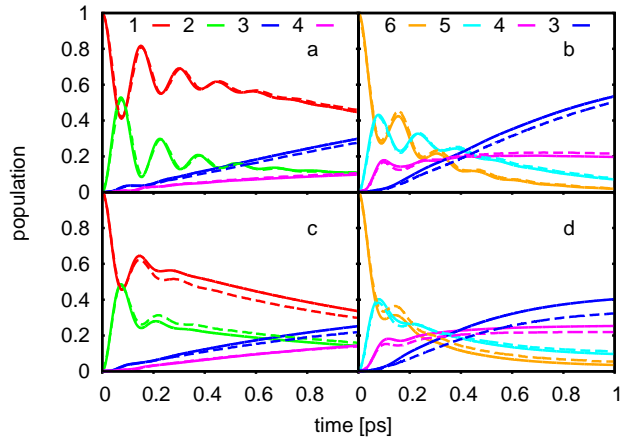}
\caption{\label{fig:transf_vs_ish_flem}
Comparison of the excitation transfer taken from Ref.~\cite{IsFl09_17255_} calculated with HEOM approach (dashed lines) and that obtained from the ZOFE master equation (solid lines).
The used spectral density is the solid curve of Fig.~\ref{fig:spec_dens}(a).
(a) Excitation probability for BChl~1 to~4 over time at temperature 77K, with initial excitation only on BChl~1.
(b) Excitation probability for BChl~3 to~6, 77K, and initial excitation only on BChl~6.
(c) As (a), but for 300K.
(d) As (b), but for 300K.
}
\end{figure}

The calculated transfer resulting from this spectral density is shown in Fig.~\ref{fig:transf_vs_ish_flem}.
The curves show the time-dependent excitation probabilities of the individual BChls --- the dashed lines are the HEOM results taken from Ref.~\cite{IsFl09_17255_} and the solid lines show the results calculated with the ZOFE master equation (see Eq.~(\ref{zofe_mast_eq})).
In the upper panels the transfer is calculated for a temperature of 77K and in the lower panels for 300K.
Initially all excitation is either on BChl~1 (left panels) or on BChl~6 (right panels).
In all these four cases there is quite good agreement between the HEOM and the ZOFE results, showing that the ZOFE approximation preserves all the detailed features of the transfer, e.g.\ the strong oscillations at 77K.

\subsection{Influence of high-energy modes}

Since the ZOFE method enables us to consider an almost arbitrary form of the spectral density,
we can now investigate the influence of different contributions to the spectral density. 

Often, the so-called reorganization energy 
\begin{equation}
  \lambda=\int_0^{\infty}d\omega\ J(\omega)/\omega
\end{equation}
is used as a global measure for the coupling strength to the (environmental) vibrational modes depending on the spectral density $J(\omega)$~\cite{ChFl09_241_}.
Then strong coupling (compared to the electronic interactions $V_{nm}$) is identified as $\lambda \gg V_{nm}$ and weak coupling as $\lambda \ll V_{nm}$.

In the following we will investigate the influence of contributions to the spectral density in different energy regions and we will show that in many cases the reorganization energy is not a reasonable measure for the coupling strength.

First let us consider the influence of the high energy part of the spectral density.
As an example we consider a spectral density (dotted curve in Fig.~\ref{fig:spec_dens}(a)), which coincides with the D-L spectral density of Ref.~\cite{IsFl09_17255_} for low energies, but has a large additional peak centered at about $1600\cm$, resulting in strong additional coupling to high energy modes.
We have calculated the transfer through the FMO monomer for this modified spectral density using the ZOFE master equation.
The resulting transfer is exactly the same as that calculated by the ZOFE method for the original spectral density of Ref.~\cite{IsFl09_17255_} shown in Fig.~\ref{fig:transf_vs_ish_flem}.
This shows that despite contributing significantly to the reorganization energy, the high energy part above $500\cm$ (up to $500\cm$ the spectral densities are the same) is of no relevance for the system dynamics.
The reason for that lies in the fact that the system energies, i.e.\ all the transition energies between the electronic eigenenergies of the FMO monomer (indicated by the vertical lines in Fig.~\ref{fig:spec_dens}), are in the energy region below $500\cm$.   
That means that the detuning between these system energies and the high energy part of the spectral density is so large (compared to the coupling strength) that the high energy modes do not couple to the system.
However, the reorganization energy does not take this fact into account and thus gives in this case a wrong indication for the strength of the coupling to the modes.

Similarly, a faster drop-off, as given by the spectral density shown by the curve (triangles) in Fig.~\ref{fig:spec_dens}(a), does not affect the transfer dynamics, since in the region below $500\cm$ it also agrees  with the original spectral density.

\subsection{Influence of low-energy modes}

The spectral densities we considered so far have a smooth form without any prominent structure.
However, a more realistic spectral density consists of several distinct peaks, that embody the strong coupling to the intra-molecular modes of the BChls, and is taken e.g.\ from measured fluorescence line narrowing spectra~\cite{AdRe06_2778_,WePuPr00_5825_} or from atomistic simulations (see e.g.~Ref.~\cite{ARXIV_Sangwoo,Unp_Olbrich}).

In the following we will investigate the influence of such a structured spectral density and the influence of how the coupling strength is shared among the peaks at different energies.  
To this end we consider a spectral density consisting of four distinct peaks.
To obtain information about the importance of the individual peaks we vary their height in a range of parameters where from previous investigations~\cite{RoStEi11_034902_} we are confident that the ZOFE approximation gives reliable results.

To distinguish the influence of the structure from that of the overall coupling strength, we want to roughly keep the total coupling strength in the relevant energy region (below $\sim 550\cm$) constant.
To this end we first introduce an effective reorganization energy  
\begin{equation}
\label{eq:eff_reorg_energ}
\lambda_{\rm eff}=\int_{E_{\rm min}}^{E_{\rm max}}d\omega\ J(\omega)/\omega,
\end{equation}
where we take $E_{\rm min}=0\cm$ and $E_{\rm max}=550\cm$ as the relevant energy range, that contains all electronic transition energies of the FMO monomer (see the vertical lines in Fig.~\ref{fig:spec_dens})\footnote{Another reasonable measure for the effective coupling strength in the relevant energy region is the integral $X_{\rm tot}=\int_{E_{\rm min}}^{E_{\rm max}}d\omega\ J(\omega)/\omega^2$ over all Huang-Rhys factors $X_{\lambda}=\kappa_{\lambda}^2/\omega_{\lambda}^2$ of the modes in this energy region. However, this integral does not converge for the special form of the considered D-L spectral density when $E_{\rm min}=0\cm$ is taken.}.

For the D-L spectral density in Fig.~\ref{fig:spec_dens}(a) we get a value of $\lambda_{\rm eff}=31\cm$ for the effective reorganization energy.

To investigate the influence of the structure of the spectral density, we consider three different spectral densities, that all have the same effective reorganization energy as the D-L spectral density.
But in contrast to the D-L spectral density, these spectral densities consist of four prominent peaks, whose intensities are varied. 
These spectral densities are shown as the solid curves in Fig.~\ref{fig:spec_dens}(b)--(d).
From Fig.~\ref{fig:spec_dens}(b) to (d), one of the peaks at a time is enhanced.
For comparison again the D-L spectral density is shown in each figure as the dotted curve.
Note, that due to the division by $\omega$ in the definition of the reorganization energy (see Eq.~(\ref{eq:eff_reorg_energ})), to keep the value of $\lambda_{\rm eff}$ constant, the peak intensity of the high-energy peaks becomes much larger than that of the low-energy peaks, when they are enhanced.

\begin{figure} 
\includegraphics[width=0.85\mylenunit]{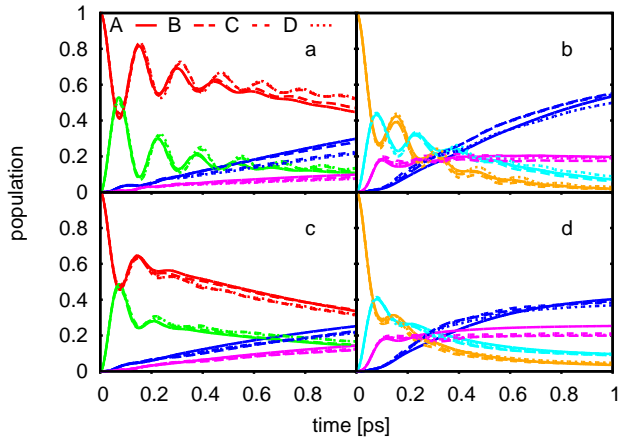}
\caption{\label{fig:transf_structured_spd}
Influence of the structure of the spectral density:
Transfer calculated with ZOFE master equation, as in Fig.~\ref{fig:transf_vs_ish_flem}, but for the three different spectral densities in Fig.~\ref{fig:spec_dens}(b)--(d) and the D-L spectral density in Fig.~\ref{fig:spec_dens}(a). 
}
\end{figure}

To investigate the influence of these changes in the spectral density, we calculated the excitation transfer through the FMO monomer using the ZOFE master equation.
The resulting time-dependent excitation probabilities are shown in Fig.~\ref{fig:transf_structured_spd} in the same way as in Fig.~\ref{fig:transf_vs_ish_flem}, but now for the three different spectral densities of Fig.~\ref{fig:spec_dens}(b)--(d) and the D-L spectral density. 

One sees that even though the overall coupling strength in the relevant energy region is kept constant, the transfer changes clearly when the structure of the spectral density is varied.
In Fig.~\ref{fig:transf_structured_spd}(a) for instance, the population on BChl~3 after $1\,ps$ changes almost by up to 50\% depending on the structure of the spectral density.

Obviously, not only the overall coupling strength in the relevant energy region is important, but also the local distribution of the coupling strength.  

When the temperature is increased (lower panels), the spreading of the curves decreases, showing that the structure of the spectral density becomes less important at higher temperatures.

\section{summary and outlook}
\label{sec:summary}

In the present paper we have discussed excitation energy transfer in the FMO complex, emphasizing the role of the coupling of the excitation to vibrational modes of the BChl molecules.
\paragraph{ZOFE master equation:}
To handle the coupling of the electronic excitation to vibrational modes the numerical calculations have been performed using the ZOFE master equation, utilizing the ZOFE approximation, which allows to calculate the transport for the  FMO complex  within a few minutes on a standard PC. 
We validated the applicability of the ZOFE approximation by comparing with the exact HEOM results of Ref.~\cite{IsFl09_17255_}, where a Drude-Lorentz spectral density has been used.

\paragraph{Reorganization energy:}
Since we are not restricted to a certain spectral density (SD), we investigated the influence of the parts of the SD in different energy regions.
As expected, the high energy part of the SD (above $\sim 500$ cm$^{-1}$) does not influence the transport properties.
In this context it is worth noting that the so-called reorganization energy, which is often used as a measure for the strength of the coupling to the environment, is often not a reasonable measure, since it takes into account the SD at all energies.
In particular SDs, that have the same reorganization energy but, in the relevant energy region, have different shapes, affect the system dynamics differently.  

\paragraph{Structured spectral density:}Being aware of this we used an ``effective reorganization energy'' (ERE), which is a local measure in the energy region of system transitions~\footnote{It has to be emphasized, that also this measure does not take details of the spectral density properly into account.}, to investigate the dependence of the transfer on the local structure of the SD.
We find that even when the structure is changed only slightly and the ERE is kept constant, we observe a change in the transfer to BChl 3 up to 50\%.
This influence of the structure of the SD decreases when temperature is increased.

\paragraph{Electronic energies and couplings:}
Note that the transfer of course depends also strongly on the precise values of the electronic couplings $V_{nm}$ between the BChls  and their  transition energies. 
For example we found that  the final population of BChl 3 increases by roughly 50\% when the electronic couplings are increased by 25\%.
Thus care has to be taken not to overestimate theoretical results obtained for a certain set of parameters.

\paragraph{Outlook:}

Since measured spectral densities  show many peaks in the  energy region where electronic transition energies are located \cite{WePuPr00_5825_}, in further work the interplay between the electronic Hamiltonian and highly structured SDs should be studied in more detail.

For such systematic studies the NMQSD (and accordingly the ZOFE master equation derived from it) seems to be well suited, due to the short calculation times.
Another big advantage of the computational efficiency of the approach is that larger systems are tractable now -- the full trimer with 24 chromophores and a structured SD can be calculated on a standard PC within a few hours.
One goal of the present study was to further investigate the rather abstract ZOFE approximation \cite{YuDiGi99_91_,RoStEi11_034902_}.
The quite good agreement  with the exact HEOM results of Ref.~\cite{IsFl09_17255_} makes us confident that it is worth to further explore the NMQSD method.

The NMQSD approach is  also  suitable to describe other light harvesting complexes and molecular aggregates.
While in the case of the FMO complex the coupling to high energy vibrational modes with energies larger than $1000 \cm$ does not influence the energy transfer markedly, in molecular aggregates of organic dyes these modes have a strong impact on the electronic excitation dynamics \cite{RoScEi09_044909_}. 
Furthermore they have energies comparable to the dipole-dipole interaction which influences strongly e.g.\  the aggregate absorption spectra \cite{EiBr06_376_}.
In previous investigations we have found that typical features of molecular aggregate spectra, such as the exchange narrowing of the J-band and the broad H-band, are reproduced by the NMQSD calculations \cite{RoStEi10_5060_}.

The NMQSD method is also readily applicable to treat optical spectroscopy.
In particular it should be well suited to investigate the influence of the coupling to vibrational modes on the signals observed in 2D experiments. 
Such a study might be revealing, since the ``power spectrum'' obtained from 2D measurements in Ref.~\cite{EnCaRe07_782_} shows strong resemblance to the fluorescence line narrowing spectrum of Wendling et.\ al \cite{WePuPr00_5825_} .

To obtain further information how individual vibrational modes influence the transport it might be useful to simulate the transport using networks consisting of quantum \cite{ARXIV_Mostame} or classical oscillators \cite{ARXIV_Briggs}.

\section{acknowledgment}
  Financial support from the DFG under Contract No. Ei 872/1-1 is acknowledged.
AE thanks Al\'{a}n~Aspuru-Guzik for the hospitality.
We thank John Briggs for helpful comments and Alexander Croy for many useful discussions and for providing the numerical code for the coth expansion of the environment correlation function.


\end{document}